\documentclass[aps,prd,twocolumn,superscriptaddress,showpacs]{revtex4}

\usepackage{graphicx}

\begin{document}

%\preprint{CDMS \today}

\title{Limits on spin-dependent WIMP-nucleon interactions \\
from the Cryogenic Dark Matter Search}


%\thanks{}

%\collaboration{CDMS Collaboration}

\author{D.S.~Akerib} \affiliation{Department of Physics, Case Western Reserve University, Cleveland, OH  44106, USA} 
\author{M.S.~Armel-Funkhouser} \affiliation{Department of Physics, University of California, Berkeley, CA 94720, USA} 
\author{M.J.~Attisha} \affiliation{Department of Physics, Brown University, Providence, RI 02912, USA} 
\author{C.N.~Bailey} \affiliation{Department of Physics, Case Western Reserve University, Cleveland, OH  44106, USA} 
\author{L.~Baudis} \affiliation{Department of Physics, University of Florida, Gainesville, FL 32611, USA} 
\author{D.A.~Bauer} \affiliation{Fermi National Accelerator Laboratory, Batavia, IL 60510, USA} 
\author{P.L.~Brink} \affiliation{Department of Physics, Stanford University, Stanford, CA 94305, USA} 
\author{P.P.~Brusov} \affiliation{Department of Physics, Case Western Reserve University, Cleveland, OH  44106, USA} 
\author{R.~Bunker} \affiliation{Department of Physics, University of California, Santa Barbara, CA 93106, USA} 
\author{B.~Cabrera} \affiliation{Department of Physics, Stanford University, Stanford, CA 94305, USA} 
\author{D.O.~Caldwell} \affiliation{Department of Physics, University of California, Santa Barbara, CA 93106, USA} 
\author{C.L.~Chang} \affiliation{Department of Physics, Stanford University, Stanford, CA 94305, USA} 
\author{J.~Cooley} \affiliation{Department of Physics, Stanford University, Stanford, CA 94305, USA} 
\author{M.B.~Crisler} \affiliation{Fermi National Accelerator Laboratory, Batavia, IL 60510, USA} 
\author{P.~Cushman} \affiliation{School of Physics \& Astronomy, University of Minnesota, Minneapolis, MN 55455, USA} 
\author{M.~Daal} \affiliation{Department of Physics, University of California, Berkeley, CA 94720, USA} 
\author{F.~DeJongh} \affiliation{Fermi National Accelerator Laboratory, Batavia, IL 60510, USA} 
\author{R.~Dixon} \affiliation{Fermi National Accelerator Laboratory, Batavia, IL 60510, USA} 
\author{M.R.~Dragowsky} \affiliation{Department of Physics, Case Western Reserve University, Cleveland, OH  44106, USA} 
\author{D.D.~Driscoll} \affiliation{Department of Physics, Case Western Reserve University, Cleveland, OH  44106, USA} 
\author{L.~Duong} \affiliation{School of Physics \& Astronomy, University of Minnesota, Minneapolis, MN 55455, USA} 
\author{R.~Ferril} \affiliation{Department of Physics, University of California, Santa Barbara, CA 93106, USA} 
\author{J.~Filippini} \affiliation{Department of Physics, University of California, Berkeley, CA 94720, USA} 
\author{R.J.~Gaitskell} \affiliation{Department of Physics, Brown University, Providence, RI 02912, USA} 
\author{S.R.~Golwala} \affiliation{Department of Physics, California Institute of Technology, Pasadena, CA 91125, USA} 
\author{D.R.~Grant} \affiliation{Department of Physics, Case Western Reserve University, Cleveland, OH  44106, USA} 
\author{R.~Hennings-Yeomans} \affiliation{Department of Physics, Case Western Reserve University, Cleveland, OH  44106, USA} 
\author{D.~Holmgren} \affiliation{Fermi National Accelerator Laboratory, Batavia, IL 60510, USA} 
\author{M.E.~Huber} \affiliation{Department of Physics, University of Colorado at Denver and Health Sciences Center, Denver, CO 80217, USA} 
\author{S.~Kamat} \affiliation{Department of Physics, Case Western Reserve University, Cleveland, OH  44106, USA} 
\author{S.~Leclercq} \affiliation{Department of Physics, University of Florida, Gainesville, FL 32611, USA} 
\author{A.~Lu} \affiliation{Department of Physics, University of California, Berkeley, CA 94720, USA} 
\author{R.~Mahapatra} \affiliation{Department of Physics, University of California, Santa Barbara, CA 93106, USA} 
\author{V.~Mandic} \affiliation{Department of Physics, University of California, Berkeley, CA 94720, USA} 
\author{P.~Meunier} \affiliation{Department of Physics, University of California, Berkeley, CA 94720, USA} 
\author{N.~Mirabolfathi} \affiliation{Department of Physics, University of California, Berkeley, CA 94720, USA} 
\author{H.~Nelson} \affiliation{Department of Physics, University of California, Santa Barbara, CA 93106, USA} 
\author{R.~Nelson} \affiliation{Department of Physics, University of California, Santa Barbara, CA 93106, USA} 
\author{R.W.~Ogburn} \affiliation{Department of Physics, Stanford University, Stanford, CA 94305, USA} 
\author{T.A.~Perera} \affiliation{Department of Physics, Case Western Reserve University, Cleveland, OH  44106, USA} 
\author{M.~Pyle} \affiliation{Department of Physics, Stanford University, Stanford, CA 94305, USA} 
\author{E.~Ramberg} \affiliation{Fermi National Accelerator Laboratory, Batavia, IL 60510, USA} 
\author{W.~Rau} \affiliation{Department of Physics, University of California, Berkeley, CA 94720, USA} 
\author{A.~Reisetter} \affiliation{School of Physics \& Astronomy, University of Minnesota, Minneapolis, MN 55455, USA} 
\author{R.R.~Ross} \thanks{Deceased} \affiliation{Lawrence Berkeley National Laboratory, Berkeley, CA 94720, USA} \affiliation{Department of Physics, University of California, Berkeley, CA 94720, USA}
\author{T.~Saab} \affiliation{Department of Physics, Stanford University, Stanford, CA 94305, USA} 
\author{B.~Sadoulet} \affiliation{Lawrence Berkeley National Laboratory, Berkeley, CA 94720, USA} \affiliation{Department of Physics, University of California, Berkeley, CA 94720, USA}
\author{J.~Sander} \affiliation{Department of Physics, University of California, Santa Barbara, CA 93106, USA} 
\author{C.~Savage} \affiliation{Department of Physics, University of California, Santa Barbara, CA 93106, USA} 
\author{R.W.~Schnee} \affiliation{Department of Physics, Case Western Reserve University, Cleveland, OH  44106, USA} 
\author{D.N.~Seitz} \affiliation{Department of Physics, University of California, Berkeley, CA 94720, USA} 
\author{B.~Serfass} \affiliation{Department of Physics, University of California, Berkeley, CA 94720, USA} 
\author{K.M.~Sundqvist} \affiliation{Department of Physics, University of California, Berkeley, CA 94720, USA} 
\author{J-P.F.~Thompson} \affiliation{Department of Physics, Brown University, Providence, RI 02912, USA} 
\author{G.~Wang} \affiliation{Department of Physics, California Institute of Technology, Pasadena, CA 91125, USA} \affiliation{Department of Physics, Case Western Reserve University, Cleveland, OH  44106, USA}
\author{S.~Yellin} \affiliation{Department of Physics, Stanford University, Stanford, CA 94305, USA} \affiliation{Department of Physics, University of California, Santa Barbara, CA 93106, USA}
\author{J.~Yoo} \affiliation{Fermi National Accelerator Laboratory, Batavia, IL 60510, USA} 
\author{B.A.~Young} \affiliation{Department of Physics, Santa Clara University, Santa Clara, CA 95053, USA} 

\collaboration{CDMS Collaboration}\noaffiliation

\date{January 11, 2006}

\begin{abstract}
The Cryogenic Dark Matter Search (CDMS) is an experiment to detect weakly interacting massive particles (WIMPs), which may constitute the universe's dark matter, based on their interactions with Ge and Si nuclei.  We report the results of an analysis of data from the first two runs of CDMS at the Soudan Underground Laboratory in terms of spin-dependent WIMP-nucleon interactions on $^{73}$Ge and $^{29}$Si.  These data exclude new regions of WIMP parameter space, including regions relevant to spin-dependent interpretations of the annual modulation signal reported by the DAMA/NaI experiment.
\end{abstract}

\pacs{95.35.+d, 14.80.Ly}

%\keywords{}

\maketitle

%\section{Introduction}
The nature of the dark matter which dominates structure formation in our universe is one of the most pressing questions of modern cosmology \cite{Primack:1988zm,Jungman:1995df,Bertone:2004pz}.  A promising class of candidates is weakly interacting massive particles (WIMPs) \cite{Steigman:1984ac}, particularly the lightest neutralino in supersymmetric (SUSY) extensions to the Standard Model~\cite{Jungman:1995df}. Many groups have sought to detect WIMPs directly via their elastic scattering off atomic nuclei \cite{Gaitskell:2004rw}.

The nucleon coupling of a slow-moving Majorana neutralino (or of any WIMP in the extreme non-relativistic limit \cite{Kurylov:2003ra}) is characterized by two terms: spin-independent (e.g.\ scalar) and spin-dependent (e.g.\ axial vector). When coherence across the nucleus is taken into account~\cite{Lewin:1995rx}, these two terms behave very differently. The neutralino has similar scalar couplings to the proton and neutron~\cite{Jungman:1995df}, and nucleon contributions interfere constructively to enhance the WIMP-nucleus elastic cross section. Thus, though neutralino-nucleon cross sections for such interactions are generally orders of magnitude \textit{smaller} than in the axial case \cite{Bednyakov:2000he}, scalar couplings dominate direct detection event rates in most SUSY models for experiments using heavy target nuclides.

In contrast, the axial couplings of nucleons with opposing spins interfere destructively, leaving WIMP scattering amplitudes determined roughly by the unpaired nucleons (if any) in the target nucleus.  Spin-dependent WIMP couplings to nuclei thus do not benefit from a significant coherent enhancement, and sensitivity to such interactions requires the use of target nuclides with unpaired neutrons or protons.  Spin-dependent interactions may nonetheless dominate direct-detection event rates in spin-sensitive experiments in regions of parameter space where the scalar coupling is strongly suppressed.  This can provide a lower bound on the total WIMP-nucleus elastic cross section, since spin-dependent amplitudes are more robust against fine cancellations \cite{Bednyakov:2004qu}.  In general, consideration of such couplings when interpreting experimental results more fully constrains WIMP parameter space and allows exploration of alternative interpretations of possible signals \cite{Bernabei:2003za,Savage:2004fn}.  In this work we explore the implications of recent results from the Cryogenic Dark Matter Search (CDMS) experiment for spin-dependent elastic scattering.  Some limits with a previous data set have appeared in \cite{Akerib:2005zy}, and constraints on spin-independent interactions with this data set are discussed in \cite{Akerib:2005R119}.

The Cryogenic Dark Matter Search \cite{Akerib:2004fq,Akerib:2005zy} seeks to detect WIMPs via their interaction with nuclei in semiconductor crystals.  CDMS uses ZIP detectors \cite{Akerib:2005zy} to discriminate between electron recoils (induced by most backgrounds) and nuclear recoils (induced by WIMPs and neutrons) on an event-by-event basis via a simultaneous measurement of ionization and athermal phonons.  Under standard assumptions about the galactic halo (described in \cite{Lewin:1995rx}), CDMS currently sets the strictest upper limits on spin-independent WIMP interactions \cite{Akerib:2004fq, Akerib:2005zy}.

The CDMS detectors are made of natural Ge or Si, both composed predominantly of spinless isotopes with negligible sensitivity to spin-dependent interactions.  However, each contains one significant isotope with non-zero nuclear spin: $^{73}$Ge (spin-9/2) makes up 7.73\% of natural Ge, while $^{29}$Si (spin-1/2) makes up 4.68\% of natural Si.  Each isotope contains a single unpaired neutron, making CDMS much more sensitive to spin-dependent interactions with neutrons than with protons.

The analysis presented here is based on the combination of two data runs taken at the Soudan Underground Laboratory, a deep installation which provides a rock overburden of 2090 meters water equivalent.  The first Soudan run used a single tower of 6 ZIP detectors (4 Ge, 2 Si) and recoil energy thresholds of 10-20 keV.  From October 2003 through January 2004, 52.6 live
days of WIMP-search data were acquired.  We work with the results of the ``current'' analysis detailed in \cite{Akerib:2005zy}, in which one candidate event was identified in Ge.  The second run added a second tower (2 Ge, 4 Si) and acquired 74.5 live days of data between March and August of 2004.  We work with the 7-keV threshold analysis described in \cite{Akerib:2005R119}, also with one candidate event in Ge.  Both candidates are consistent with expected backgrounds, and no candidates were observed in Si.  Scaling the exposures before analysis cuts by the isotopic abundances given above, we obtain a total of 11.5 (1.7) raw kg-days $^{73}$Ge ($^{29}$Si) exposure.

\begin{figure*}
\includegraphics[width=8cm]{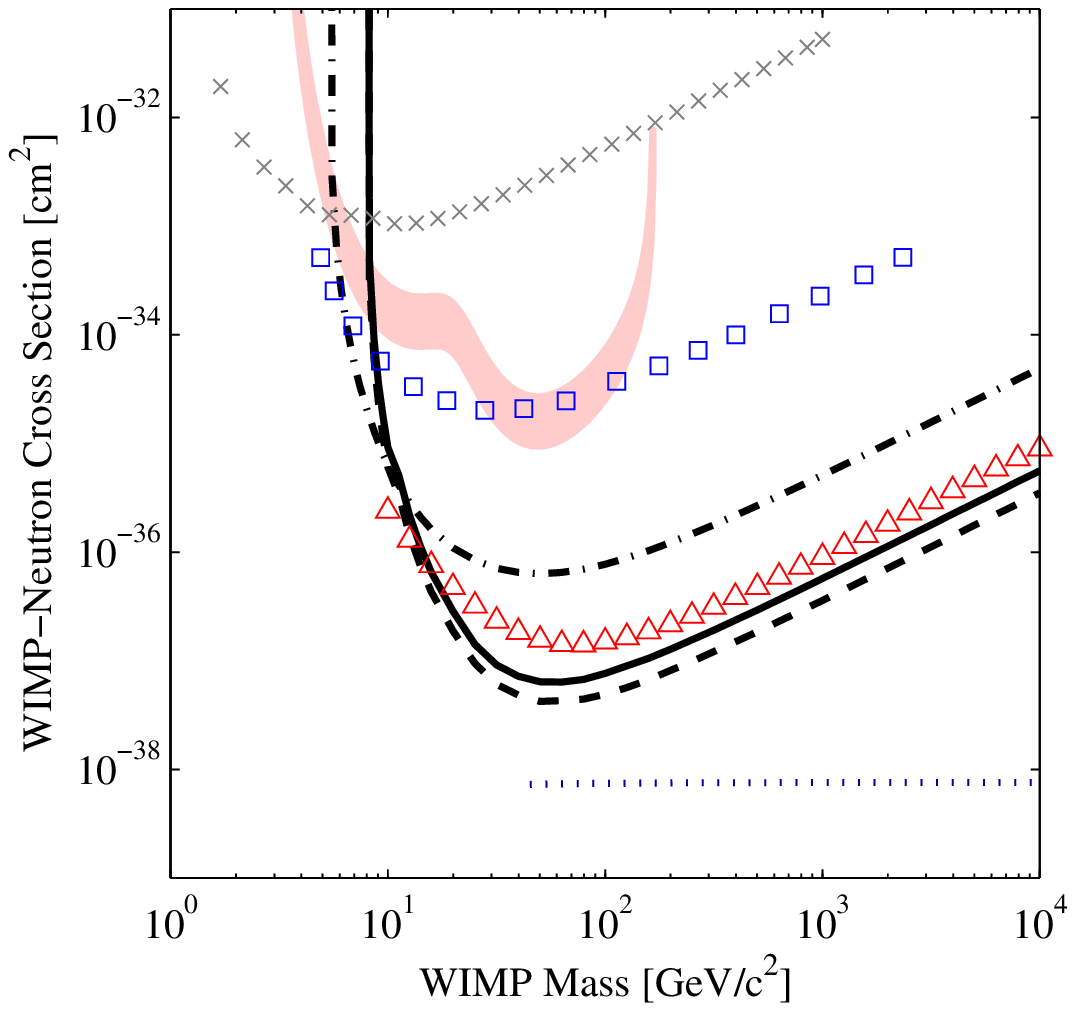}\includegraphics[width=8cm]{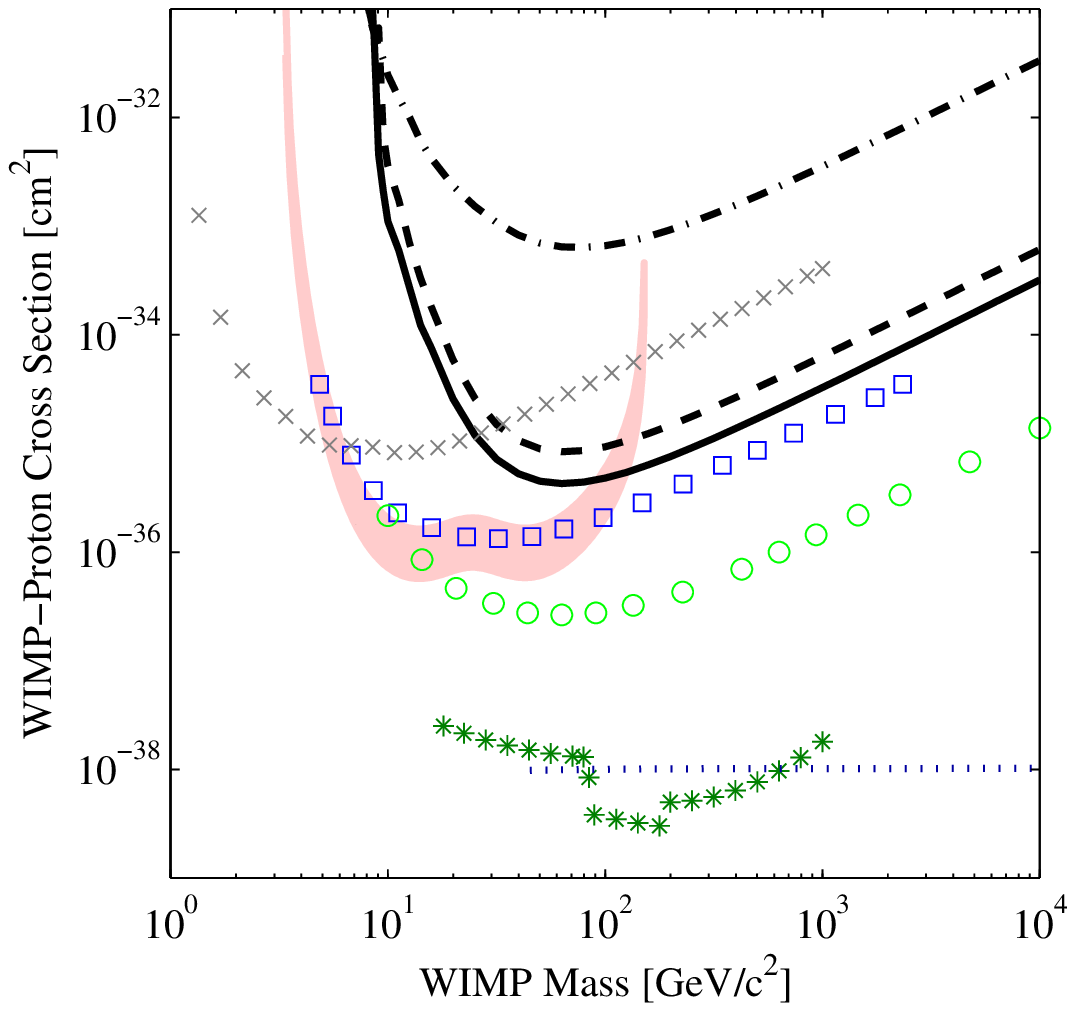}
	\caption{\label{fig:sigmaNP}Upper limit contours (90\% confidence level) for recent CDMS data sets, plotted in the cases of (\textit{left}) pure neutron and (\textit{right}) pure proton coupling.  We show limits based on Ge (solid) and Si (dash-dot) from the combined Soudan data (black). Dashed curves represent Ge limits using the alternate form factor from \cite{Ressell:1993qm}. As benchmarks, we also include interpretations of the DAMA/NaI annual modulation signal \cite{Bernabei:2003za} (filled regions are $3\sigma$-allowed) and limits from other leading experiments: CRESST I \cite{Angloher:2002in} as computed in \cite{Savage:2004fn} (``$\times$''s), PICASSO \cite{Barnabe-Heider:2005pg} (squares), NAIAD \cite{Alner:2005kt} (circles), ZEPLIN I \cite{Kudryavtsev:2004sd} (triangles), and Super-Kamiokande \cite{Desai:2004pq} (asterisks; an indirect search, based on different assumptions).  EDELWEISS \cite{Benoit:2004tt} and SIMPLE \cite{Girard:2005pt} report limits (not shown for clarity) comparable to CDMS Si and PICASSO, respectively.  As a theoretical benchmark (see text), horizontal dotted lines indicate expected cross sections for a heavy Majorana neutrino.  Plots courtesy of \cite{Gaitskell:dmplotter}.}
\end{figure*}

The spin-dependent interactions of a given WIMP are characterized by its WIMP-proton and WIMP-neutron spin-dependent couplings, $a_p$ and $a_n$.  Unlike in the spin-independent case, the relative strengths of these couplings may vary significantly with neutalino composition \cite{Tovey:2000mm}, though they are often of similar magnitude in models of interest \cite{Bednyakov:2003wf}. The differential cross section for spin-dependent WIMP-nucleus elastic scattering at momentum transfer $q$ can be written as~\cite{Engel:1991wq}
\begin{equation}\label{eqn:sigmaSD}\frac{d\sigma^{SD}_{\chi N}}{dq^2}=\frac{8G_F^2}{(2J+1)v^2}S(q), \end{equation}
where $v$ is the incident WIMP velocity, $J$ is the nuclear spin, $G_F$ is Fermi's constant, and the ``spin structure function'' $S(q)$ is given by
\begin{equation} S(q)=a_0^2S_{00}(q)+a_0a_1S_{01}(q)+a_1^2S_{11}(q), \end{equation}
with $a_0=a_p+a_n$ and $a_1=a_p-a_n$.  The functions $S_{ij}(q)$ encompass the magnitude of the spin associated with the nucleon populations, as well as the effects of the spatial distribution of that spin at non-zero momentum-transfers.  These must be determined separately for each nuclide using a nuclear structure model.  Such models may be compared based on their $q=0$ nucleon spin expectation values ($\langle S_p \rangle$ and $\langle S_n \rangle$) and the accuracy of their predictions for the nuclear magnetic moment.  These models also show that spin correlations (e.g. polarization of the even nucleon group by the odd nucleon group) give a nucleus with an unpaired neutron a residual sensitivity to $a_p$.

For $^{29}$Si, the major efforts to determine nuclear spin structure have been large-basis shell model simulations by Ressell et al. \cite{Ressell:1993qm} and Divari et al. \cite{Divari:2000dc}.  The results of both calculations agree in the zero-momentum-transfer limit and reproduce the experimental magnetic moment ($\mu=-0.555\mu_N$) to within $10\%$.  We follow the former.

The most complete shell model studies of the high-spin $^{73}$Ge nuclide have been carried out by Ressell et al. \cite{Ressell:1993qm} and Dimitrov et al. \cite{Dimitrov:1994gc}.  The former result requires ``quenching'' to bring its predicted value of the nuclear magnetic moment ($\mu=-1.239\mu_N$) in line with experiment ($\mu=-0.879\mu_N$), while the hybrid model used in the latter does not ($\mu=-0.920\mu_N$).  Both models give values of $\langle S_n \rangle$ within $\sim2$\% of one another, but their values for $\langle S_p \rangle$ differ by a factor of 3.   We follow Dimitrov et al., but also compute Ge limits following Ressell et al.\ (the ``alternate form factor'') to give an indication of nuclear model uncertainties.  Note that the structure-function fits given in the latter are invalid for recoil energies $E_R>50$ keV.  We thus assume no sensitivity to such recoils when using this model (thereby limiting our sensitivity for high WIMP masses in this case).

We follow a ``model-independent'' framework described in \cite{Tovey:2000mm,Giuliani:2004uk, Savage:2004fn} for the interpretation of experimental results in terms of spin-dependent interactions.  For easy comparisons between experiments, we report allowed regions in two $M_{\chi}-\sigma^{SD}$ planes: one in the limit $a_n=0$, one for $a_p=0$.  For this purpose, we express limits in terms of WIMP-nucleon cross sections
\begin{equation}\label{eqn:Toveysigma}\sigma^{SD}_{p,n} = \frac{8 (J+1)}{\pi J} G_F^2 \mu_{\chi p,n}^2 a_{p,n}^2, \end{equation}
where $\mu_{\chi p,n}$ is the WIMP-nucleon reduced mass.
We also plot allowed regions in the $a_p-a_n$ plane for two choices of WIMP mass, $M_{\chi}$.

\begin{figure*}	\includegraphics[width=8cm]{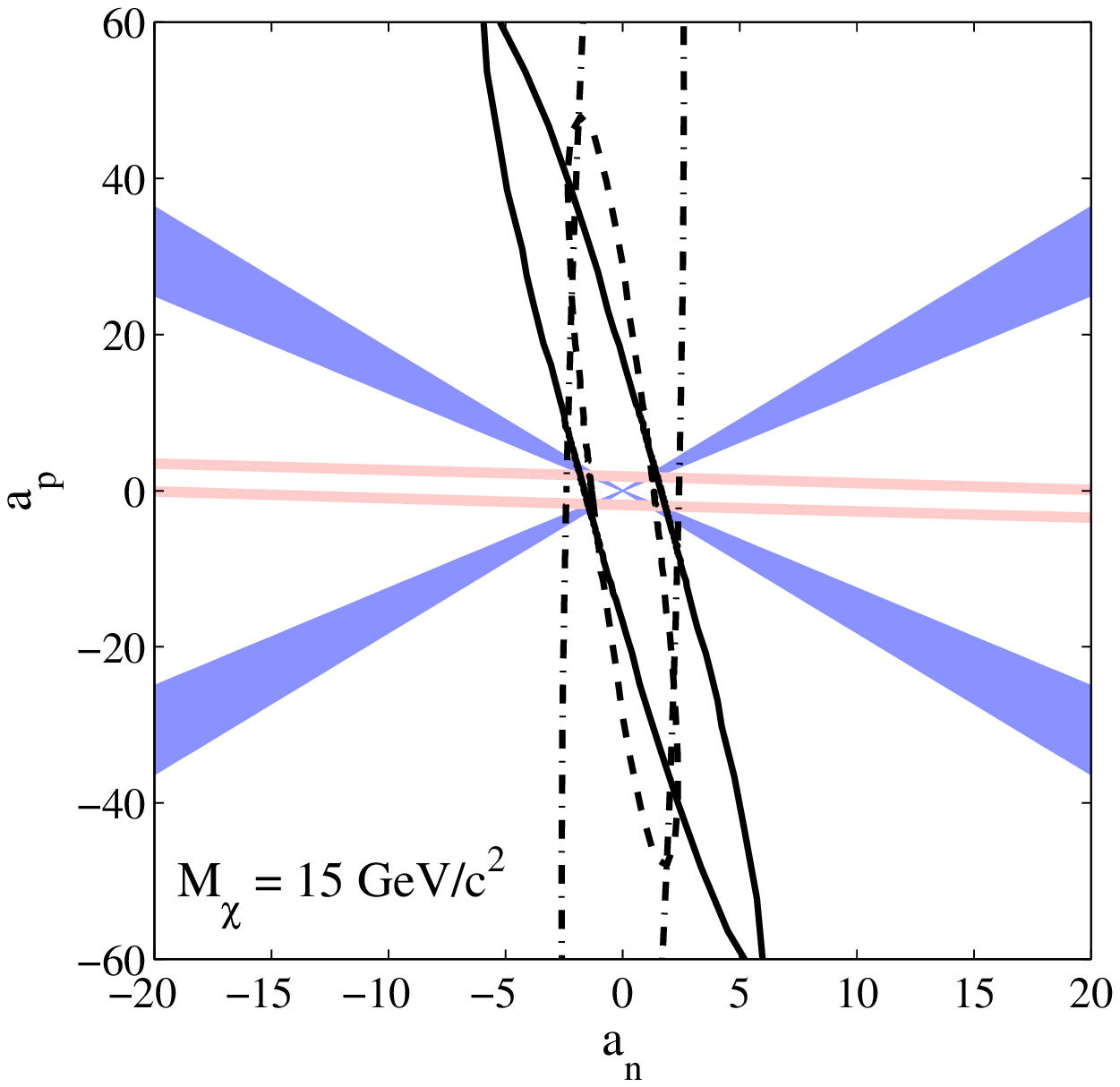} \includegraphics[width=8cm]{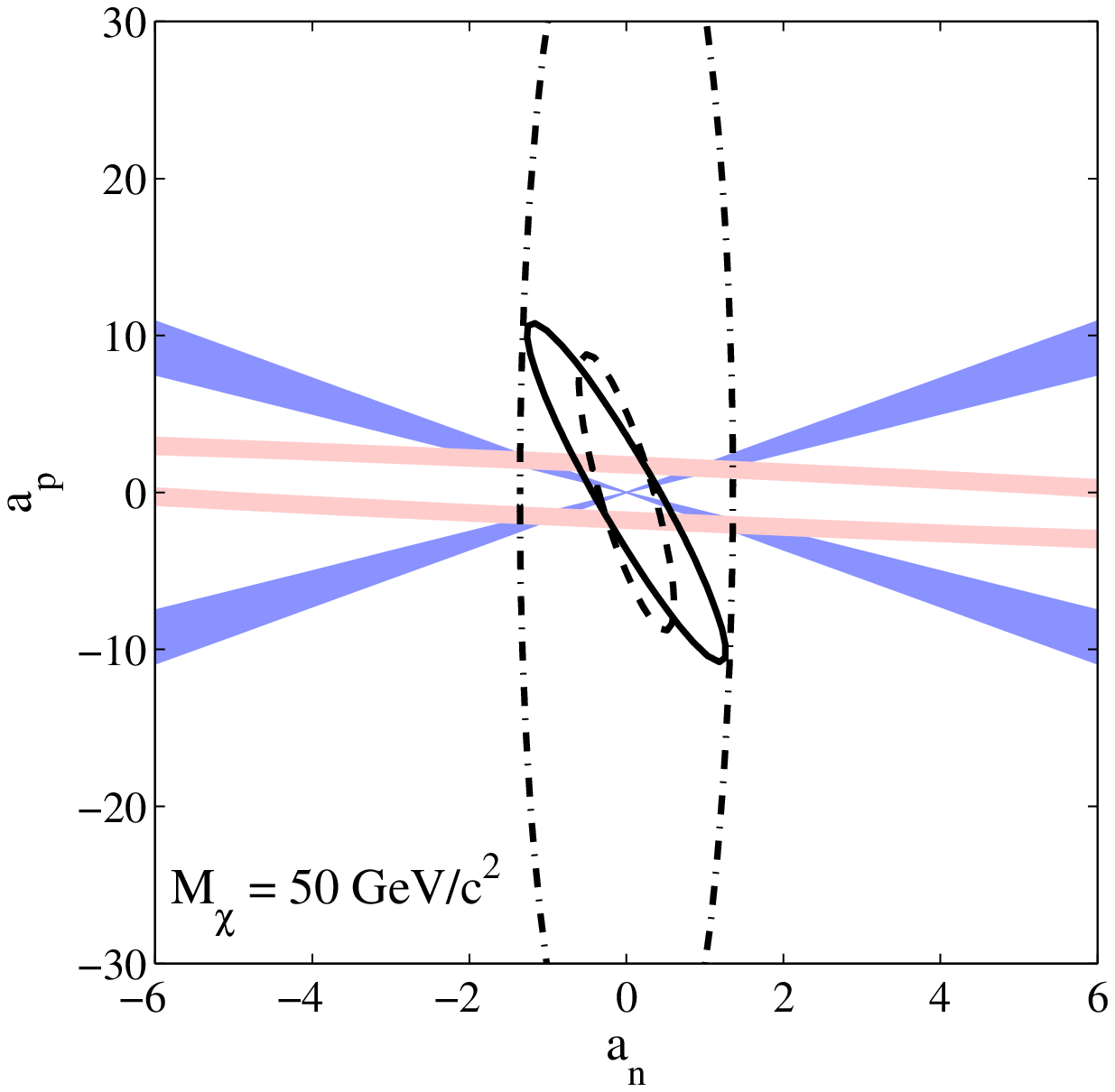}
	\caption{\label{fig:apan2050}Regions in the $a_p-a_n$ plane allowed (at the 90\% confidence level) by CDMS data.  Each data set excludes the exterior of the corresponding ellipse.  Two choices of WIMP mass $M_{\chi}$ are shown: 15\,GeV/c$^2$ (\textit{left}) and 50\,GeV/c$^2$ (\textit{right}).  Dot-dashed ellipses represent Si limits, solid ellipses represent Ge limits, and dashed ellipses represent Ge limits using the alternate form factor.  Also shown are the corresponding interpretations of the DAMA/NaI modulation signal (near-horizontal light (pink) filled bands are $3\sigma$-allowed).  The thin dark (blue) filled wedges correspond to models satisfying $0.55<|\frac{a_n}{a_p}|<0.8$, a constraint from the effMSSM framework \cite{Bednyakov:2003wf}.}
\end{figure*}

Figure~\ref{fig:sigmaNP} shows upper-limit contours in the $M_{\chi}-\sigma^{SD}$ plane in the limiting cases of pure neutron coupling ($a_p=0$) and pure proton coupling ($a_n=0$) for the CDMS data sets, computed using Yellin's Optimum Interval method \cite{Yellin:2002xd}.  Dashed curves in Fig.~\ref{fig:sigmaNP} and \ref{fig:apan2050} show the limits obtained using the alternate form factor.  As benchmarks, we also include recent limits from other leading experiments, as well as $3\sigma$ allowed regions based on DAMA/NaI's reported annual modulation \cite{Bernabei:2003za}.  The latter is computed following \cite{Savage:2004fn}, based on the 2--4~keVee modulation amplitude alone and a tripling of its quoted 1$\sigma$ error bars.  Including constraints from other energy bins favors somewhat lower cross sections at high WIMP masses, but does not modify our conclusions.

Due to its isotopic composition, CDMS is sensitive primarily to WIMP-neutron spin-dependent couplings.  The combined CDMS data exclude new regions of parameter space in the case of purely WIMP-neutron coupling (i.e. for $a_p=0$).  In combination with CRESST I \cite{Angloher:2002in}, these data are inconsistent with an interpretation of the DAMA/NaI annual modulation amplitude in terms of such interactions within the standard halo model (see also \cite{Savage:2004fn}).  Despite its lack of unpaired protons, CDMS also possesses competitive sensitivity to $a_p$.  CDMS does not currently set the strongest limits in the pure WIMP-proton case, but has begun to explore the region of this parameter space associated with the DAMA/NaI signal.

To explore more general models, these results can also be expressed in the $a_p-a_n$ plane for various choices of WIMP mass.  Two such choices (15\,GeV/c$^2$ and 50\,GeV/c$^2$) are shown in Fig.~\ref{fig:apan2050}.  The former illustrates the advantage of using two active isotopes: the region allowed by both nuclides (approximately the overlap of the corresponding ellipses) may be significantly smaller than either ellipse individually.  In the current case, the generally weaker limit from $^{29}$Si serves to cut off the regions at large $a_p^2+a_n^2$ allowed by Ge.  This is significant since the lengths of the major axes of these ellipses depend on near-cancellations in the structure functions, and so have substantial uncertainties.  Other experiments (see \cite{Savage:2004fn}) set more stringent constraints on $a_p$ and further reduce the overall allowed region.

Specific WIMP model frameworks yield constraints on the relationship between $a_p$ and $a_n$.  In particular, SUSY neutralinos are expected to have comparable couplings to protons and neutrons.  As an example, Bednyakov \cite{Bednyakov:2003wf} finds limits of $0.55<|\frac{a_n}{a_p}|<0.8$ in the effMSSM framework, corresponding to the thin filled wedges in Fig.~\ref{fig:apan2050}.  Imposing this constraint eliminates the overlap between the CDMS and DAMA/NaI allowed regions at WIMP masses $\gtrsim25$\,GeV/c$^2$, though compatible regions at lower masses remain.

The upper limits set here do not yet constrain SUSY models significantly.  An increase in exposure by at least two orders of magnitude is required to explore the most accessible mSUGRA models \cite{Baltz:2005}, which may be possible with the next generation of direct detection experiments.

One benchmark for spin-sensitive dark matter searches is a heavy Majorana neutrino, which has a purely axial coupling to nucleons.  Horizontal dotted lines in Fig.~\ref{fig:sigmaNP} indicate the expected spin-dependent cross sections of such a particle \cite{Lewin:1995rx}.  These lines suggest the magnitude of spin-dependent cross section expected from a Majorana dark matter candidate with weak interactions (regardless of production mechanism), and thereby indicate an approximate upper bound to ``interesting'' WIMP parameter space.  We should note that this particle is intended as a benchmark rather than a serious dark matter candidate.  Such a neutrino could be unstable, and relic density calculations suggest that it cannot be a major component of the dark matter if produced thermally~\cite{Srednicki:1988ce}.  CDMS II will begin to probe this parameter space in the near future, and some has already been reached by indirect searches.

\begin{acknowledgments}
The authors thank E. Baltz and J. Engel for useful conversations, and the technical and support staff at our various institutions.  This work is supported by the National Science Foundation under Grant No. AST-9978911 and No. PHY-9722414, by the Department of Energy under contracts DE-AC03-76SF00098, DE-FG03-90ER40569, DE-FG03-91ER40618, and by Fermilab, operated by the Universities Research Association, Inc., under Contract No. DE-AC02-76CH03000 with the Department of Energy. The ZIP detectors were fabricated in the Stanford Nanofabrication Facility of NNIN supported by NSF under Grant ECS-9731293.
\end{acknowledgments}

%\bibliography{PRDr118SD}

\end{document}